\documentclass[aps,nofootinbib,preprint,floatfix]{revtex4}

\usepackage{graphicx}
\usepackage{amsmath}
\usepackage{float}
\usepackage{microtype}	
\usepackage[toc,page]{appendix}
\usepackage{subfigure}
\usepackage{multirow}
\usepackage{pdflscape}
\usepackage{alphalph}

\setcitestyle{square,super}
\newcommand*{\citen}[1]{%
  \begingroup
    \romannumeral-`\x 
    \setcitestyle{numbers}%
    \cite{#1}%
  \endgroup
}

\makeatletter
\newalphalph{\alphmult}[mult]{\@alph}{26}
\makeatother
\renewcommand{\subsection}{\alphmult{\value{subsection}}}

\setcitestyle{square,super}

\usepackage{color}   
\usepackage{hyperref}
\newcommand{\red}{\textcolor{black}}
\hypersetup{
    colorlinks=true, 
    linktoc=all,     
    linkcolor=blue,  
}

\begin{document}

\title{The importance of basis states: an example using the Hydrogen basis}
\author{Lindsay Forestell and Frank Marsiglio}
\affiliation{Department of Physics, University of Alberta, Edmonton, Alberta, Canada, T6G~2E1}

\begin{abstract}

We use a simple system, the electron configuration in a Hydrogen-like atom, to demonstrate the importance of 
using a complete basis set to provide a proper quantum mechanical description. We first start with what might
be considered a successful strategy --- to diagonalize a truncated Hamiltonian matrix, written in a basis consisting
of Hydrogen ($Z=1$) basis states. This fails to provide the correct answer, and we then demonstrate that the
continuum basis states provided the rest of the true wave function, for the bound ground states. This work then
shows, in a relatively simple system, the need to utilize a complete basis set, consisting of both bound and continuum
states.
\end{abstract}

\date{\today} 
\maketitle


\section{Introduction}
Undergraduate texts on Quantum Mechanics generally emphasize analytical solutions to the Schr\"odinger Equation. But
they almost always include a section on Hilbert Space, and basis states, and the formal solution of a problem through
matrix mechanics. In a number of recent papers, \cite{marsiglio09,jelic12,jugdutt13} we have illustrated, with familiar
examples (Harmonic oscillator, Coulomb, etc.), the determination of bound state energies and eigenfunctions through a matrix formulation of quantum mechanics. All of these examples have used the familiar basis states which are the eigenstates
for the infinite square well. In all the cases considered, we have truncated the Hilbert Space to a manageable size, so that
the low-lying eigenvalues and their accompanying eigenfunctions are determined to high accuracy.

At the same time, while not really emphasized in textbooks, students are generally made aware that a proper Hilbert
Space needs to complete, to be useful as a basis set. This message was driven home recently in a study of the Helium
electronic ground state, utilizing a basis set consisting of two-electron product states of Hydrogenic bound state
eigenfunctions. \cite{hutchinson13} While this is the natural basis set to use to understand Helium (and, indeed, any
multi-electron atom of the periodic table), Hutchinson et al. \cite{hutchinson13} emphasized that these product states
constitute an incomplete basis set, and an accurate description requires the continuum states as well. Quantum chemists
learned this lesson long ago,\cite{hylleraas29} and therefore never use such a (technically) poor basis set. As
argued in Ref. \citen{hutchinson13}, however, for state-of-the-art research problems on correlated electron
systems, sometimes one does not have a choice.

The case of the Helium electronic ground state is sufficiently complicated that the lesson in Ref. \citen{hutchinson13}
may be beyond the reach of undergraduates. Our purpose is to present a much simpler case, that
of the one electron Hydrogen-like atom with central charge $Z_0$, which we call `$Z_0{\rm Hydrogen}$'. Here of 
course the energies and states are a trivial extension to Hydrogen itself, where the Bohr radius $a_0$ is simply changed 
to $a_0/Z_0$. 

We start first by supposing that we do not know this answer, but instead decided to formulate the problem as a matrix
problem, using as a basis the familiar bound states of the Hydrogen atom. There are an infinite number of these bound states,
so we must necessarily truncate. As shown in the next section the result converges very rapidly, and no more than ten
or so Hydrogen bound states are required to attain a converged result, but it is the wrong result ! The reason is explained, also
in physical terms, and in the ensuing section we make use of the known exact ground state for this problem to project out
the contributions from each Hydrogenic state, including the continuum ones. These coefficients now sum to unity, as should
be the case, for a proper description of $Z_0{\rm Hydrogen}$. \red{We also illustrate how the continuum wave functions with increasing
momentum contribute to the final (correct) wave function.}

\section{Attempt at a Matrix Formulation for $Z_0{\bf Hydrogen}$}

We are interested in the ground state solution for the problem of a single electron interacting with a fixed nucleus of
charge $Z_0$. This has a Hamiltonian given by
\begin{equation}
H = \frac{-\hbar^2}{2m}\nabla^2 - Z_0 \frac{e^2}{4\pi \epsilon_0} \frac{1}{r}
\label{z0_ham}
\end{equation}
\noindent where $\hbar$ is Planck's constant, $m$ is the mass of the electron, $-e$ is the charge of the electron,
$eZ_0$ is the charge of the nucleus, $\epsilon_0$ is the permittivity of free space, and
$r \equiv |\vec{r}|$ is the radial coordinate. The exact eigensolutions to this problem
acquire the usual three quantum numbers, $(n,\ell,m)$, and as for the case of Hydrogen, the ground state has $n=1$ and
$\ell = m = 0$, and has energy $E^{(Z_0)}_0 = -Z_0^2E^{(0)}_1$, where 
$E^{(0)}_1 \equiv \hbar^2/(2ma_0^2) \equiv 13.6 {\rm eV}$, and
$a_0 \equiv 4\pi \epsilon_0 \hbar^2/(me^2)$ is the Bohr radius. We have assumed the nucleus to be infinitely heavy 
so the same mass appears in Eq. (\ref{z0_ham}) as appears in the definition of the Bohr radius. The exact ground state 
is given by\cite{griffiths05,levine09}
\begin{equation}
\psi^{(Z_0)}_{100} = {\red{2} \over \sqrt{4\pi}} \biggl({Z_0 \over a_0}\biggr)^{3/2} e^{-Z_0r/a_0},
\label{z0_groundstate}
\end{equation}
\noindent where the $1/\sqrt{4\pi}$ is the contribution from the angular part of the wave function for all $s$-states.

Now if we proceed as described in the Introduction, pretending not to have knowledge of this result, we would first
expand the (unknown) ground state wave function in terms of a `handy' set of basis states, which we denote as $\phi_i$,
with the index in principle representing multiple quantum numbers. Since we are using a central potential, and we
anticipate the ground state to have s-wave symmetry, then all the basis states have this symmetry as well. Thus,
if we use the Hydrogen bound states as a basis set, then the label `$n$' will denote the principal quantum number
$n$, and $\ell = m = 0$. Writing
\begin{equation}
| \psi \rangle = \sum_{n=1}^\infty c_n | \phi_n \rangle,
\label{expand}
\end{equation}
\noindent and taking inner products with each (orthonormal) basis state and the Schr\"odinger Equation written in this basis 
results in\cite{remark1} the matrix equation
\begin{equation}
\sum_{m=1}^\infty H_{nm} c_m = Ec_m.
\label{expandb}
\end{equation}
\noindent This represents an infinite dimensional matrix equation, so to make progress we truncate at $n_{\rm max}$,
vary this maximum number, and monitor the convergence of the ground state energy, for example. The required matrix
elements are
\begin{equation}
H_{nm} = \langle \phi_n |H|\phi_m\rangle.
\label{matele}
\end{equation}
The simplest way to proceed is to rewrite the Hamiltonian, Eq. (\ref{z0_ham}), as
\begin{equation}
\begin{aligned}
H &= \frac{-\hbar^2}{2m}\nabla^2 - Z_0 \frac{e^2}{4\pi \epsilon_0} \frac{1}{r} \\
&= \frac{-\hbar^2}{2m}\nabla^2 
- \frac{e^2}{4\pi \epsilon_0}\frac{1}{r} 
- (Z_0-1) \frac{e^2}{4\pi \epsilon_0}\frac{1}{r} \\
&= H_0 + H^\prime,
\end{aligned}
\label{ham}
\end{equation}
\noindent where $H_0$ is the actual Hamiltonian for Hydrogen (first two terms in second line) and $H^\prime$ is
the remaining term.
We begin with the diagonal terms, $\langle \phi_n | H | \phi_n \rangle$. Because $\phi_n$ is an eigenstate of the Hydrogen hamiltonian, these terms simplify drastically:
\begin{equation}
\begin{aligned}
\langle \phi_n | H_0+H' | \phi_n \rangle &= E_n+\langle \phi_n |H' | \phi_n \rangle \\
&= E_n - (Z_0-1) \frac{e^2}{4\pi \epsilon_0}\langle \phi_n |\frac{1}{r} | \phi_n \rangle \\
&= -\frac{E_1^{(0)}}{n^2}-(Z_0-1)\frac{e^2}{4\pi \epsilon_0}\frac{1}{n^2 a_0}, \\
&= E_1^{(0)}\frac{1}{n^2}(1-2Z_0).
\end{aligned}
\label{diag}
\end{equation}
Here we have used the results for the energy levels of hydrogen ($E_n = -\frac{E_1^{(0)}}{n^2} $), and the well known 
result\cite{griffiths05} $\langle \phi_n |\frac{1}{r} | \phi_n \rangle = 1/(n^2 a_0)$.

Because the individual $\phi_n$ are orthonormal eigenstates of $H_0$, the off-diagonal terms reduce to a simple inner product,
\begin{equation}
\begin{aligned}
\langle \phi_n | H_0 + H' | \phi_m \rangle &= 
E_n\langle \phi_n | \phi_m \rangle + \langle \phi_n | H' | \phi_m \rangle \\
&= - (Z_0-1) \frac{e^2}{4\pi \epsilon_0}\langle \phi_n |\frac{1}{r} | \phi_m \rangle \\
&= -2(Z_0-1)E_1^{(0)}\langle \phi_n |\frac{a_0}{r} | \phi_m \rangle.
\end{aligned}
\label{off_diag}
\end{equation}

Knowing the eigenstates for Hydrogen, these integrals are individually straightforward. A general formulation requires
the wavefunctions of hydrogen:\cite{remark2}
\begin{equation}
\begin{aligned}
\phi_{nlm} &= \sqrt{\biggl(\frac{2}{na_0}\biggr)^3\frac{(n-l-1)!}{(2n)[(n+l)!]^3}}
e^{-\frac{r}{na_0}}\biggl(\frac{2r}{na_0}\biggr)^l\biggl[L^{2l+1}_{n-l-1}\biggl(\frac{2r}{na_0}\biggr)\biggr]Y^m_l(\theta,\phi) \\
\phi_{n00} &=
\sqrt{\biggl(\frac{2}{na_0}\biggr)^3\frac{(n-1)!}{(2n)[(n)!]^3}}
e^{-\frac{r}{na_0}}\biggl[L^{1}_{n-1}\biggl(\frac{2r}{na_0}\biggr)\biggr]\frac{1}{\sqrt{4\pi}},
\end{aligned}
\label{hydrogen_wavefunction}
\end{equation}
\noindent where in the second line $\ell = m = 0$.
These are then substituted into the inner product in Eq.~(\ref{off_diag}). The angular integral can be done immediately, as there is no angular dependence, and this eliminates the $\frac{1}{\sqrt{4\pi}}$; furthermore, note that hereafter the letter `m' denotes
a principal quantum number (not the azimuthal quantum number which is now always zero):
\begin{equation}
\begin{aligned}
\langle \phi_m | \frac{a_0}{r} |\phi_n \rangle &= 
\frac{4}{a_0^2} \sqrt{\frac{(n-1)!(m-1)!}{n^4m^4(n!m!)^3}}\int_o^{\infty}r^2 dr \frac{1}{r} e^{-\frac{r}{na_0}}e^{-\frac{r}{ma_0}}L^1_{n-1}\biggl(\frac{2r}{na_0}\biggr)L^1_{m-1}\biggl(\frac{2r}{ma_0}\biggr) \\
&= \frac{4}{(nm)^{\frac{5}{2}}}\frac{1}{n!m!}
\int_0^\infty dx x e^{-x\frac{n+m}{nm}}L^1_{n-1}\biggl(\frac{2x}{n}\biggr)L^1_{m-1}\biggl(\frac{2x}{m}\biggr)
\\
&= \frac{4}{(nm)^{\frac{5}{2}}}\frac{1}{n!m!} I_{nm},
\end{aligned}
\label{number}
\end{equation}
\noindent where $I_{nm}$ is simply a number. This integral can be done numerically. Alternatively,
an analytic solution to the integral is achieved\cite{schrodinger26} by writing out the associated Laguerre polynomials 
as a (finite) power series.\cite{remark3} Using this series makes the required integral elementary, so we end up with
\begin{equation}
\begin{aligned}
\langle \phi_m | \frac{a_0}{r} |\phi_n \rangle = 
4\frac{(mn)^\frac{1}{2}}{(m+n)^2} ~~&\sum_{i=0}^{m-1}
\frac{(-1)^i}{i!} \frac{(m-1)!}{(m-1-i)!(i+1)!}\biggl(\frac{2n}{m+n}\biggr)^i
\\
& \sum_{j=0}^{n-1}
\frac{(-1)^j}{j!} \frac{(n-1)!}{(n-1-j)!(j+1)!}\biggl(\frac{2m}{m+n}\biggr)^j
(1+i+j)!
\end{aligned}
\end{equation}
\noindent which is simply a number. Thus, the 
off-diagonal results for the Hamiltonian are:
\begin{equation}
\langle \phi_m | H | \phi_n \rangle = -2(Z-1)E_1^{(0)} \langle \phi_m | \frac{a_0}{r} |\phi_n \rangle.
\end{equation}

\begin{figure}[h!]
\begin{center}
\includegraphics[width=0.45\columnwidth]{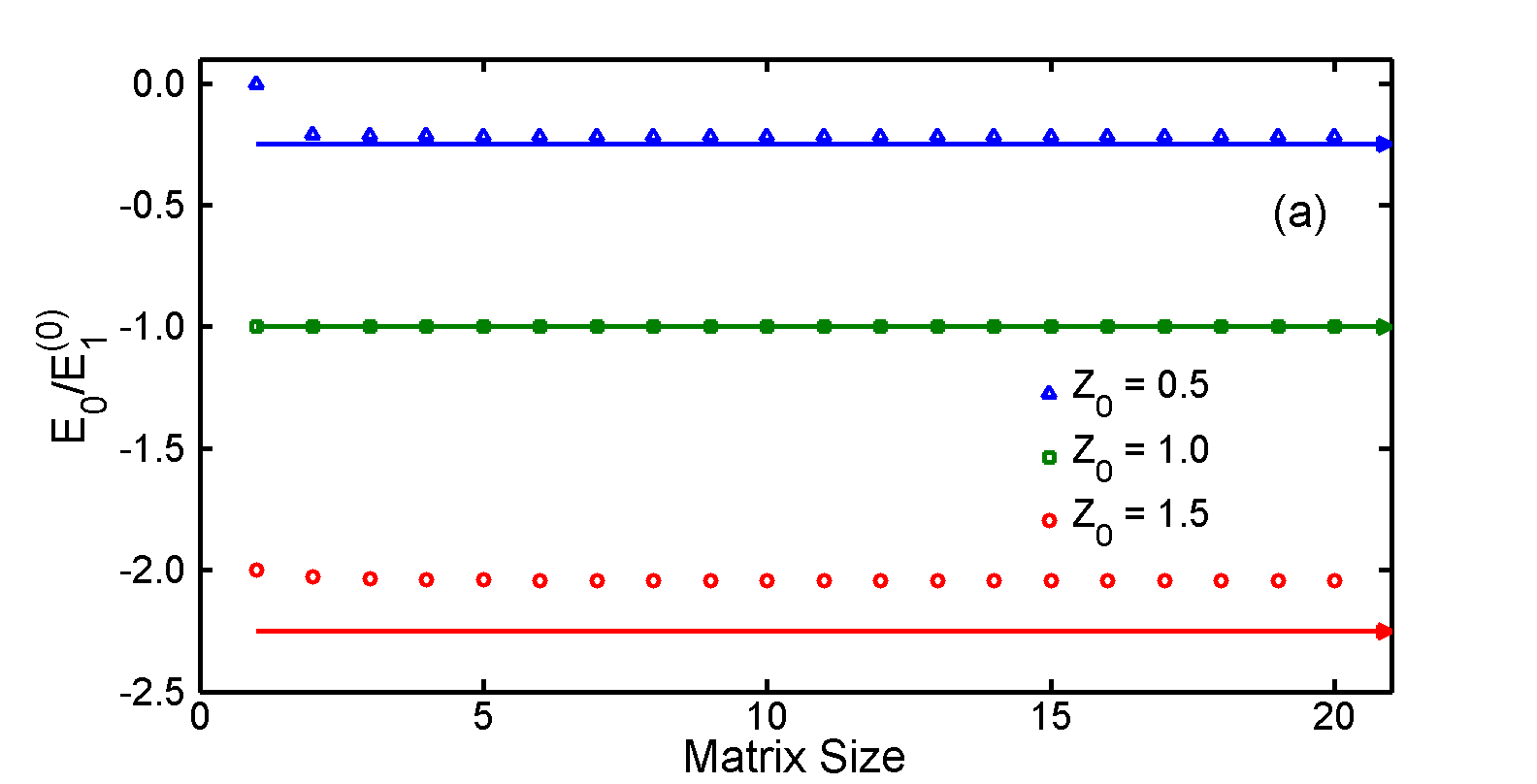}
\includegraphics[width=0.45\columnwidth]{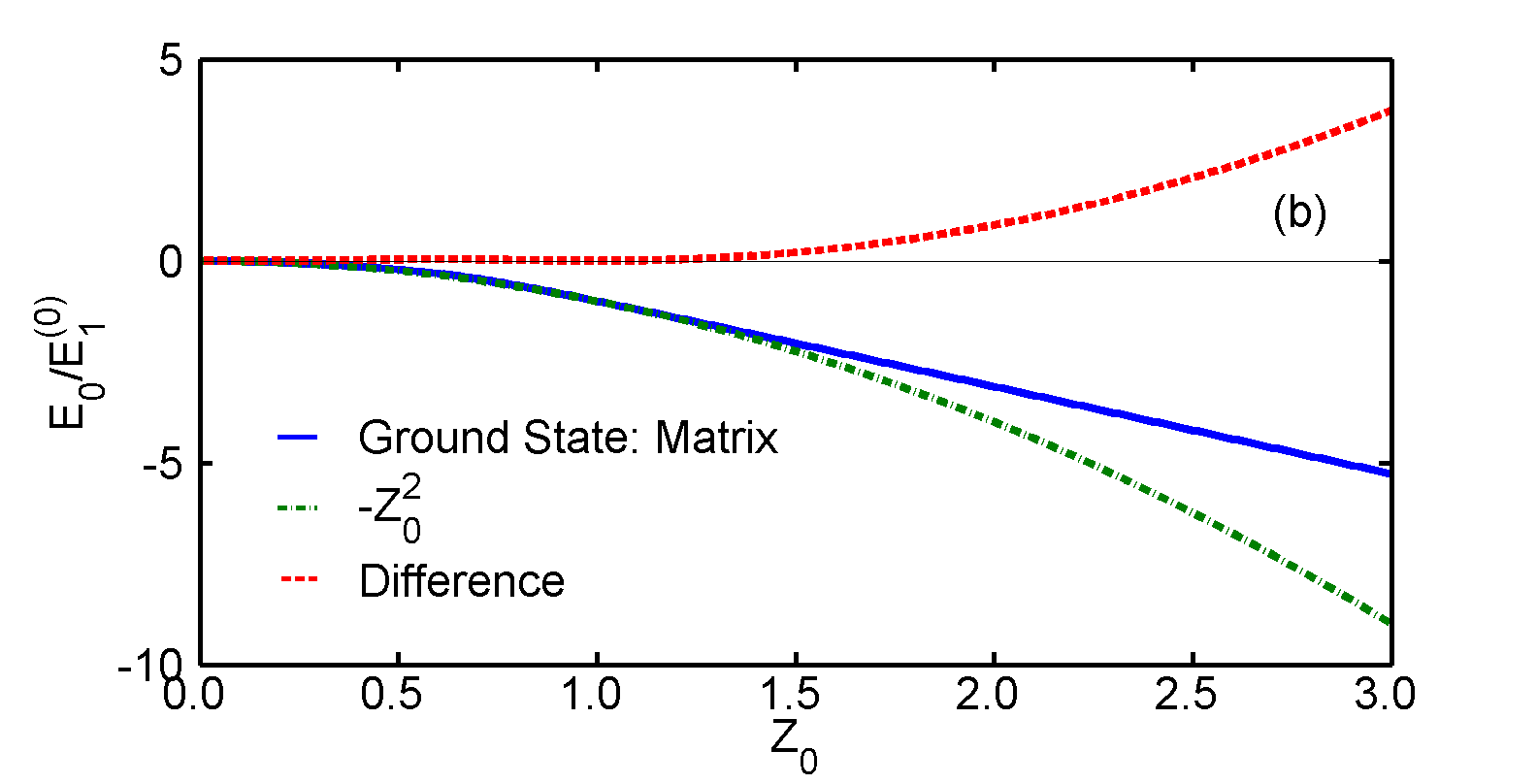}
\includegraphics[width=0.45\columnwidth]{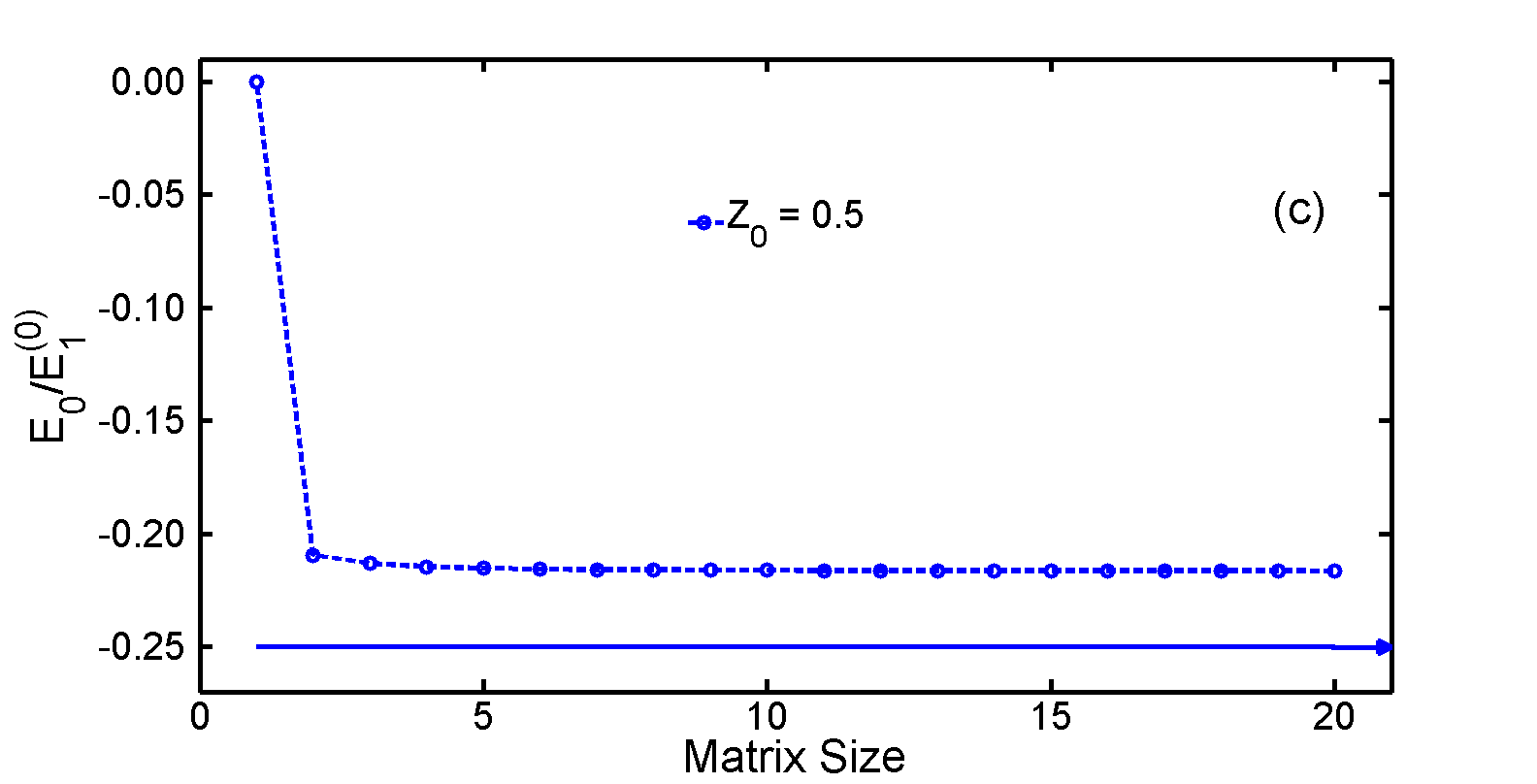}
\includegraphics[width=0.45\columnwidth]{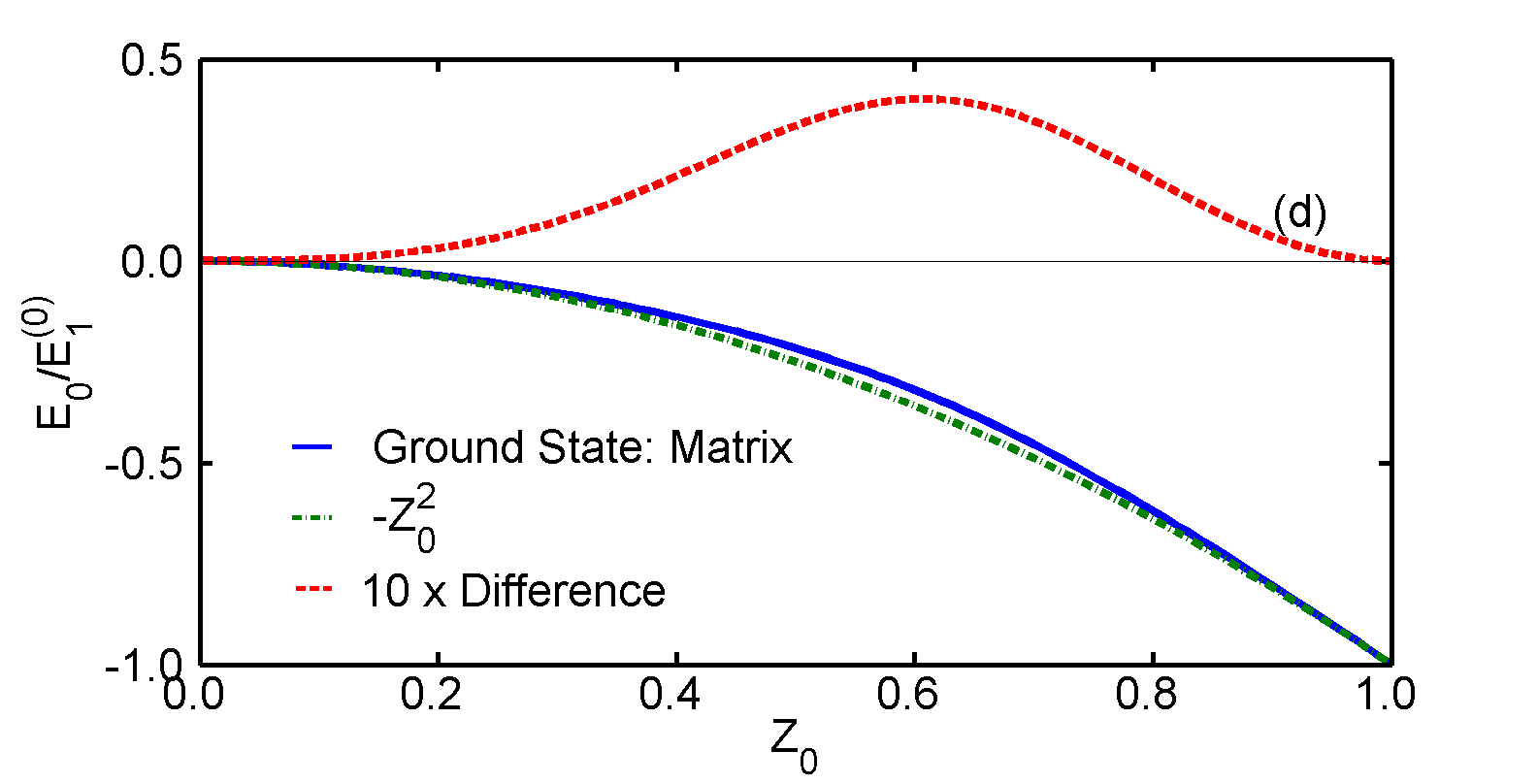}

\caption{(color online) (a) Ground state energy (in units of Rydbergs) for various values of $Z_0$ as a function
of matrix size. Note that in all cases convergence is achieved with $n_{\rm max} \approx 5$. 
(b) Ground state energy (in units of Rydbergs)  vs. $Z_0$ (solid blue curve) using $n_{\rm max} = 18$. Also
shown is the exact result, $E_0/E_1^{(0)} = -Z_0^2$ (green dashed curve) and the difference (dotted
red curve). The deviation grows as $Z_0$ increases, but notice some small discrepancies for $Z_0 < 1$ as well.
(c) An expanded view of the $Z_0 = 0.5$ result from part (a), to better illustrate the discrepancy with the exact result.
(d) An expanded view of the dependence on $Z_0$, for $Z_0 < 1$, showing that the difference peaks near
$Z_0 \approx 0.6$. Note that $10 \times$ the difference is plotted to see it better.}
\label{fig1}
\end{center}
\end{figure}

Following the philosophy of Ref. \citen{marsiglio09}, we can simply determine the Hamiltonian matrix up to
some maximum cutoff, $n_{\rm max}$, and diagonalize it to determine the ground state energy.
The results are shown in Fig. \ref{fig1}. In Fig.~(\ref{fig1}a)  we see that as we increase the size of the matrix for different values of $Z_0$, the ground state energy does converge (almost immediately). However, the energies converge to the wrong 
value (except the case of $Z_0=1$, which is simply hydrogen, and obviously needs only the one basis state for the correct
answer). The expected value is also indicated - it is simply $-Z_0^2 E_1^{(0)}$, as noted previously. In Fig.~(\ref{fig1}b)
we show the actual energy achieved vs. $Z_0$, along with the exact result, and their difference. These clearly diverge,
especially as $Z_0$ increases. 


What went wrong? As already mentioned in the Introduction, the bound state eigenstates for Hydrogen, while infinite in 
number, do not actually form a complete basis set. We should not have expected to get the correct result. On the other hand, truncating the Hilbert space has worked in previous studies of one and three dimensional problems.\cite{marsiglio09,jugdutt13}
The problem here is that as we increase $Z_0$, we are trying to describe an atom whose electron is more tightly bound
than in Hydrogen (note that $\langle r \rangle \propto a_0/Z_0)$. But utilizing the bound excited states attempts to make
use of states that are {\it more extended}, not less extended. Note that for $Z_0 < 1$ the calculation is more accurate. However,
in Fig.~(\ref{fig1}c) we show an expanded view of the result for $Z_0 = 0.5$, and even here the matrix result disagrees with the exact
result by a small but definite amount. A closer examination of the entire low $Z_0$ region is provided in Fig.~(\ref{fig1}d)
and we see that the error has a maximum
about halfway between $Z_0 = 1$ and $Z_0=0$. The relative success in this region is because physically we are trying to
construct a state that is more extended compared to Hydrogen, and the excited states are helpful in producing this. In contrast,
for $Z_0 > 1$ we require the continuum states as well, as they are perfectly capable of describing a more closely
bound state (recall that an infinite set of plane waves can describe a $\delta$-function). We demonstrate this in the next section.

\red{It should be noted that for the sake of pedagogy a different tact could have been followed. One could imagine wanting to solve for
the ground state of $Z_0{\rm Hydrogen}$, not in terms of the bound eigenstates of Hydrogen, but in terms of Hydrogenic states
with nuclear charge $Z_{\rm ref}$, where the subscript `ref' is short for `reference'. This is more artificial then the problem already considered, but it
is instructive to consider anyways, and we do so in the Appendix.}

\section{The contribution from the continuum and a full spectral decomposition}

Once we recognize the need to include the continuum states in the correct description, it immediately becomes
difficult to formulate this problem as a finite-sized matrix diagonalization problem. Here, however, we know the
exact solution, so it is still possible to determine the degree to which each basis state contributes to the overall
solution; this will clearly vary with $Z_0$. With the radial part of the wave function denoted by
$R_{10}^{Z_0}(r) = \sqrt{4\pi} \psi_{100}^{Z_0}(r)$ [see Eq. (\ref{z0_groundstate})], the correct spectral 
decomposition is given by
\begin{equation}
R_{10}^{Z_0}(r) = \sum_{n=1}^\infty a_{n}R_{n0}(r) +\int dp a_p R_{p0}(r),
\label{corr}
\end{equation}
\noindent where the $a_{n}$ and $a_p$ coefficients can be determined by overlap integrals (assuming
the left-hand-side is known), and these refer to the discrete ($n$) and continuum ($p$) components, respectively.
Note that in both cases symmetry considerations require only $\ell = 0$.

The procedure will be most familiar for the discrete coefficients, so we begin with these. We take
overlap integrals of Hydrogenic wave functions (with $Z=1$) with the exact
wave function, i.e.
\begin{equation}
a_n = \int_0^\infty \ dr \ r^2 2 \biggl({Z_0 \over a_0}\biggr)^{3/2} e^{-Z_0 r/a_0} R_{n0}(r),
\label{abound}
\end{equation}
\noindent The radial wave function $R_{n0}(r)$ can be written in terms of  an associated 
Laguerre polynomial, which has a known finite polynomial expansion;\cite{remark3} then 
the integral in Eq. (\ref{abound}) can be readily done, and what remains are 
finite sums which one can recognize as binomial expansions. The final result is
\begin{equation}
a_n = {8 Z_0^{3/2} \over (1 + Z_0)^3} \delta_{n,1} + (Z_0 - 1) (1 - \delta_{n,1}) {8n^{5/2} Z_0^{3/2} \over (1+nZ_0)^4} \biggl( {nZ_0 - 1\over nZ_0 + 1} \biggr)^{n-2}.
\label{abound2}
\end{equation}
\noindent Note the explicit factor of $(Z_0 -1)$ in front of the second term; for $Z_0 = 1$, 
the only non-zero coefficient is the $n=1$ term, with coefficient unity, as 
must be the case. For $Z_0 \ne 1$, all other $s$-states contribute with 
diminishing amplitude as $n$ increases.

A numerical summation of the probabilities $|a_n|^2$ over all values of $n$ 
reveals that these do not sum to unity (when $Z_0 \ne 1$), i.e. a finite contribution must come from 
the continuum states. The continuum states for $\ell = 0$ are denoted $R_{p 0}(r)$, where
$p$ is a continuum momentum. These eigenstates for the Hydrogen atom are less familiar to
students, but they are given in standard undergraduate texts:\cite{landau65,merzbacher98}
\begin{equation}
R_p(r) = {Z\over a_0} \sqrt{2 \pi {p a_0 \over Z} \over 1 - e^{-2\pi {Z \over pa_0}}} 
e^{-ipr} M(1 + i{Z \over pa_0}, 2, 2ipr),
\label{cont_radial}
\end{equation}
where we have written this for general $Z$ but require only $Z=1$, and
\begin{equation}
M(a,b,z) \equiv \sum_{m=0}^\infty {(a)_m \over (b)_m}{z^m \over m!}
\label{kummer}
\end{equation}
is the so-called Kummer function,\cite{abramowitz72,nist10} and 
$(a)_m \equiv a(a+1)(a+2)...(a+m-1)$ is the so-called Pochhammer
symbol. Note that $(a)_0 \equiv 1$, and $(2)_m = (m+1)!$, for example. 
Equation (\ref{cont_radial}) is a general solution, and is connected to the 
Laguerre polynomials that describe the 
radial bound state wave functions.\cite{anal}

\begin{figure}[h!]
\begin{center}
\includegraphics[height=0.8\columnwidth, angle=-90]{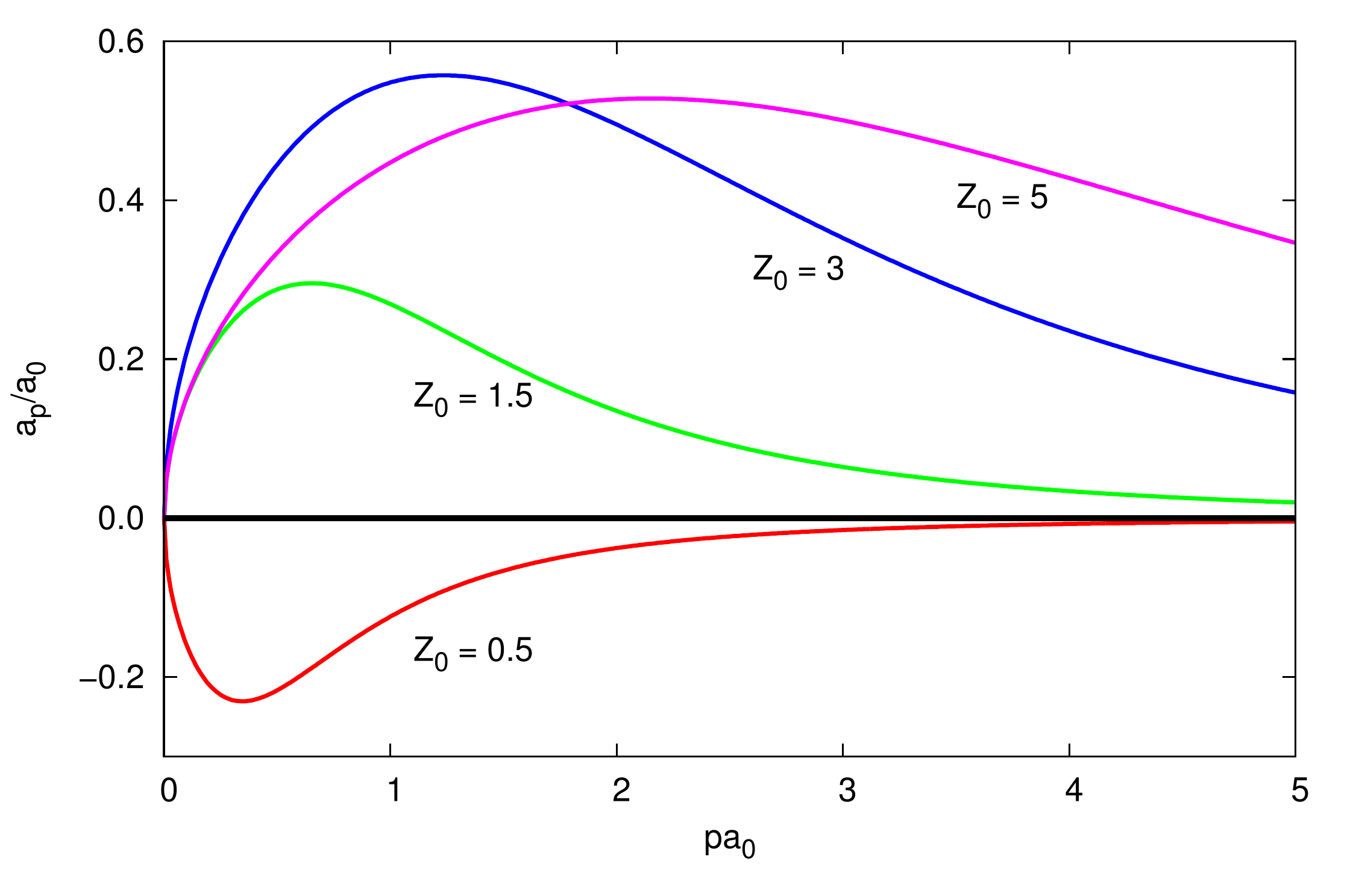} 
\caption{(color online)
\red{The momentum distribution $a_p/a_0$ given in Eq.~(\ref{ap}) as a function of $pa_0$ for various values of $Z_0$.
For $Z_0 < 1$ there is a negative correction, i.e. the continuum states try to make the approximation based on the bound states
more extended, while for $Z_0 > 1$ the corrections are all positive, so try to increase the value of the wave function near the origin.
The scale on which the continuum momentum eigenstates contribute most increases with increasing $Z_0$.}}
\label{fig2}
\end{center}
\end{figure}

\begin{figure}[h!]
\begin{center}
\includegraphics[height=0.8\columnwidth, angle=-90]{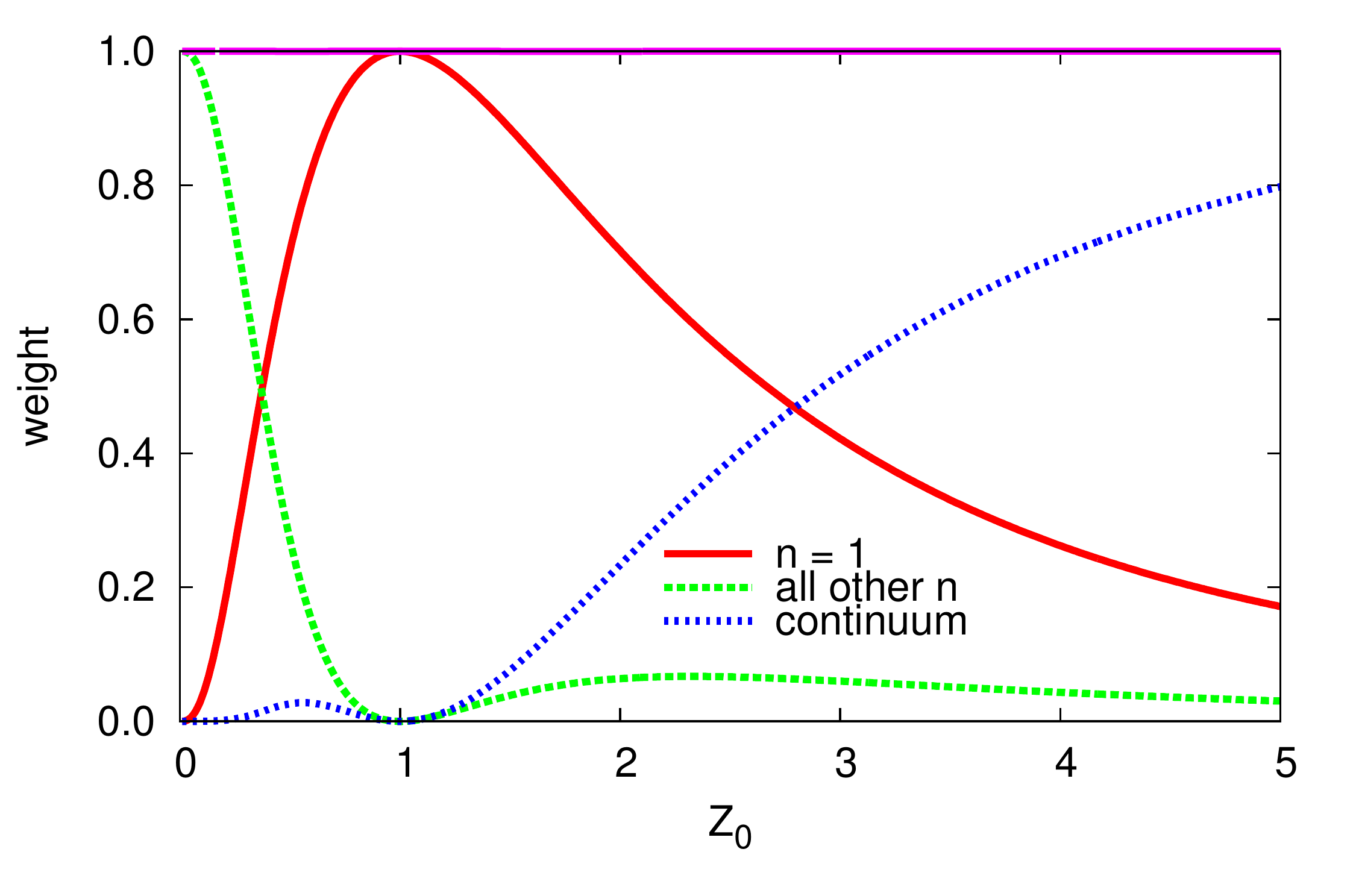} 
\caption{(color online)
Various contributions from the Hydrogenic bound and continuum states to the
ground state wave function for a central charge with magnitude $eZ_0$, as a function of
$Z_0$. At $Z_0 = 1$ only the $1s$ state contributes, as expected. For $Z_0 \ne 0$ all
other states contribute as well. In particular, for $Z_0 > 1$ the continuum states contribute
with increasing amplitude for reasons explained in the text. The sum of the three curves is
unity for all $Z_0$, \red{as indicated by the thick horizontal (purple) line across the top.}}
\label{fig3}
\end{center}
\end{figure}

The standard\cite{landau65} normalization condition for the 
continuum states,
\begin{equation}
\int_0^\infty \ dr \ r^2 R_{p^\prime}(r) R_{p}(r) = \delta (p^\prime - p),
\label{norm}
\end{equation}
determines the coefficient in Eq. (\ref{cont_radial}). By utilizing the expansion in 
Eq. (\ref{kummer}) we can evaluate the overlap integral required to obtain the 
coefficient $a_p$, for $Z=1$,
\begin{equation}
a_p = {4 \over \sqrt{a_0}}{(Z_0a_0)^{3/2} (Z_0 - 1) \over [ (pa_0)^2 + Z_0^2]^2}\sqrt{2 \pi pa_0 \over
1 - e^{-{2 \pi \over pa_0}}} \exp{\biggl[-{2 \over pa_0} {\rm tan}^{-1}\bigl( {pa_0 \over Z_0}  \bigr)\biggr]}.
\label{ap}
\end{equation}
These coefficients, when squared, show a distribution peaked around 
$p \approx Z_0/a_0$, as expected. 
\red{Since these enter linearly in the wave function expansion given in Eq.~(\ref{corr}),
we show in Fig.~(\ref{fig2}) the amplitudes, $a_p/a_0$ vs. $pa_0$ for a variety of values of $Z_0$,
showing how contributions from the continuum peak near $pa_0 \approx Z_0$. The larger $Z_0$, the larger
is the contribution from the continuum states.}

To sum the continuum contributions 
one requires a factor of $2/\pi$ to account for the enumeration of
continuum states; then the contribution from the all the continuum states is
\begin{equation}
P_{\rm cont} = {2 \over \pi} 32 \pi {(Z_0 - 1)^2 \over Z_0^3} \int_0^\infty dy 
{y \over (y^2 + 1)^4} {1 \over 1 - e^{-2 \pi \over yZ_0}}
\exp{\bigl(-{4 \over yZ_0} {\rm tan}^{-1}y\bigr)}.
\label{pcont}
\end{equation}

\begin{figure}[h!]
\begin{center}
\includegraphics[height=0.8\columnwidth, angle=-90]{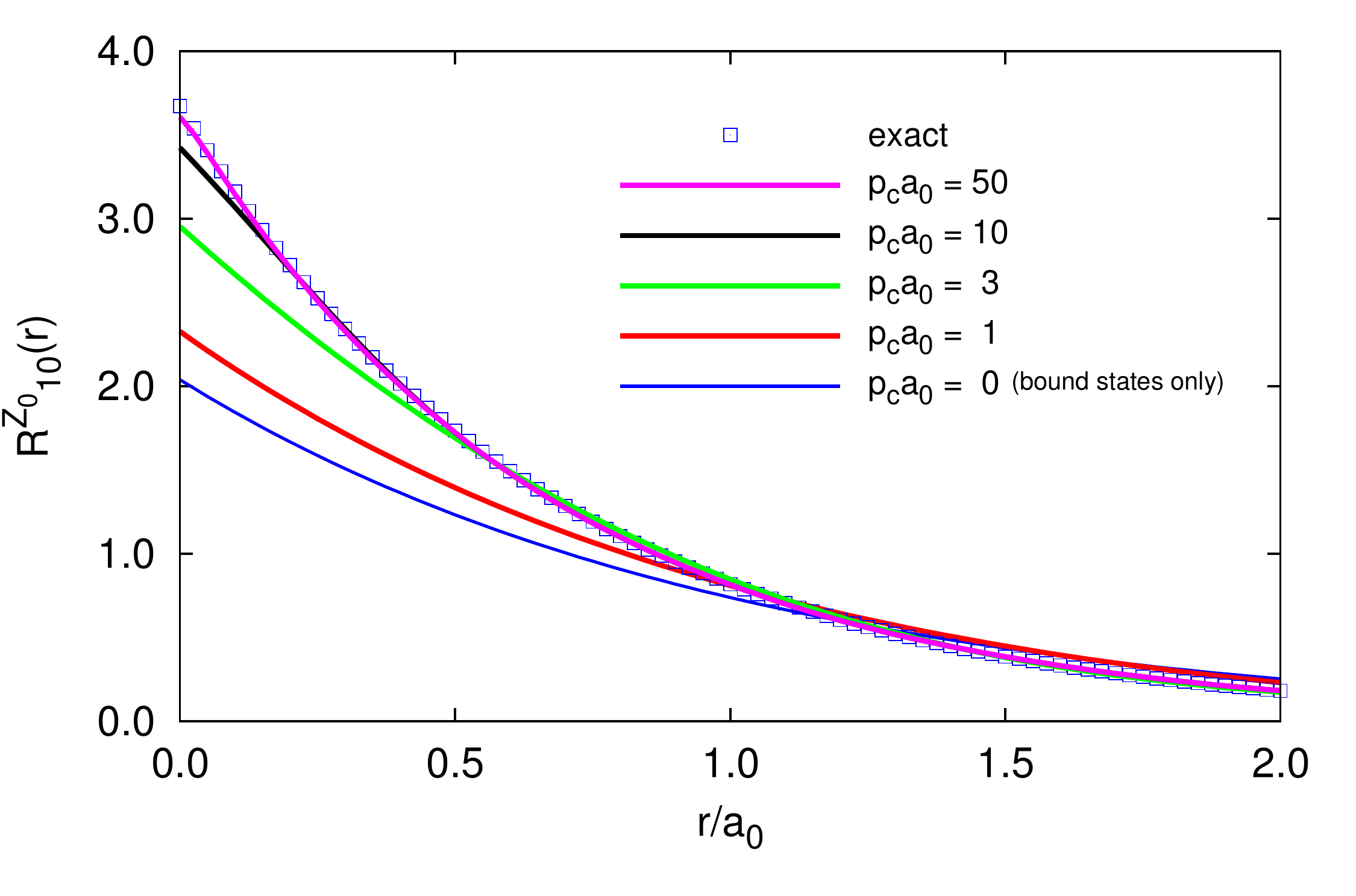} 
\caption{(color online)
\red{The evolution of the ground state wave function from just considering the bound states, i.e. the lowest (blue) curve
to the exact result (blue squares). As the accuracy increases we cut off the momentum integration in Eq.~(\ref{corr})
at $p_ca_0$ = 1 (next highest (red) curve), $p_ca_0$ = 3 (next highest (green) curve),  $p_ca_0$ = 10 (next highest (black) curve), 
and finally, $p_ca_0$ = 50 (next highest (red) curve). The higher momentum continuum components contribute to the
wave function near the origin.}}
\label{fig4}
\end{center}
\end{figure}

The contributions from the various states to the ground state are displayed in Fig.~(\ref{fig3}). We
have checked that for all $Z_0$ the contributions sum to unity. Note that for $Z_0>1$
the continuum states play an increasingly important role. Let us repeat the reason, now that we Fig.~(\ref{fig3}).
In this regime the actual wave function, given by Eq. (\ref{z0_groundstate}), varies on a scale of $a_0/Z_0$. For
$Z_0>1$ none of the bound states can provide structure on such a scale; as $n$ increases
the Hydrogenic basis states become more extended, not less. The only source of 
variation on this scale is the spectrum of continuum states, and these are readily utilized.

\red{The degree to which different continuum momentum eigenstates contribute is illustrated in Fig.~(\ref{fig4}),
where we show the wave function given by Eq.~(\ref{corr}) for a particular example, $Z_0 = 1.5$, 
but with the momentum integration cut off at increasing values of momentum. As compared with the exact wave function
it is clear that initial components (near $pa_0 \approx 0$) first fix the large $r$ behavior (the bound states produced a
value that was too high for $r/a_0 {{ \atop >} \atop {\sim \atop }} 1$) and components with larger value of momentum then
begin to produce larger amplitude near the origin. This figure demonstrates explicitly what we were just explaining, that
finer scale contributions were necessarily produced by the continuum states.}

This example serves to show how a `poor' basis choice (i.e. using the Hydrogenic set for $Z=1$
rather than for $Z=Z_0$) can require the use of the continuum states. For the case of a
single electron bound to a positive charge, a `poor' choice can obviously be avoided. However, in the
case of many electron problems where locally one can have either a single electron or two
electrons, it is very difficult to devise a basis set that diagonalizes both scenarios.\cite{hutchinson13}

\section{Summary}

We have used one of the simplest yet realistic quantum systems, the electron configuration in
a Hydrogen-like atom, to demonstrate a principle with which most students are familiar, but likely few have
encountered in practice: the need for a complete basis set to describe properly a quantum mechanical system.
The Hydrogen-like atom serves as a good system to show this, since we know the exact solution, and hence
know in advance the correct answers. We can also do most of the required integrals, both for the matrix formulation,
and for the spectral decomposition. The matrix formulation did {\em not} work, perhaps counter to what some might
expect. The implication of this failure, that continuum states are necessary, was reinforced by carrying out a spectal
decomposition. This result, portrayed in Fig.~(\ref{fig3}), perhaps runs counter to our intuition, that bound states require
a partial amplitude (that can be substantial!) corresponding to continuum states, i.e. states in which the 
particle is free to roam through all space. We also provided a physical understanding of why these states were 
especially required when we attempt to construct states that are {\em more tightly} bound than provided by the 
bound basis functions, \red{as demonstrated explicitly in Fig.~(\ref{fig4}).}

\begin{acknowledgments}

We thank Don Page and Bob Teshima for helpful discussions concerning the treatment of the continuum states. This work was supported in part by the Natural Sciences and Engineering Research Council of Canada (NSERC), by the Alberta
iCiNano program, and by a University of Alberta Teaching and Learning Enhancement Fund (TLEF) grant. 

\end{acknowledgments}

\appendix
\section{\red{A Matrix formulation using $Z_{\rm ref} \ne 1$}}

\red{As in Section [II]  we wish to solve for the Hamiltonian
\begin{equation}
H = \frac{-\hbar^2}{2m}\nabla^2 - Z_0 \frac{e^2}{4\pi \epsilon_0} \frac{1}{r},
\label{ham_zreff}
\end{equation}
but we now rewrite this using a slightly different decomposition:
\begin{equation}
\begin{aligned}
H &= \frac{-\hbar^2}{2m}\nabla^2 
- \frac{e^2}{4\pi \epsilon_0}\frac{Z_{\rm ref}}{r} 
+ (Z_{\rm ref} - Z_0) \frac{e^2}{4\pi \epsilon_0}\frac{1}{r} \\
&= H_0 + H^\prime.
\end{aligned}
\label{a_ham}
\end{equation}
\noindent The motivation is that we need more tightly bound wave functions to describe ultimately the 
ground state for $Z_0{\rm Hydrogen}$. But if $Z_{\rm ref} >> Z_0$ then many of the bound states will actually
be more `compact' compared to the ground state we are seeking to find. Therefore, proceeding as before,
we begin with the diagonal terms, $\langle \phi_n | H | \phi_n \rangle$, to obtain
\begin{equation}
\begin{aligned}
\langle \phi_n | H_0+H' | \phi_n \rangle &= E_n+\langle \phi_n |H' | \phi_n \rangle \\
&= E_n + (Z_{\rm ref} - Z_0) \frac{e^2}{4\pi \epsilon_0}\langle \phi_n |\frac{1}{r} | \phi_n \rangle \\
&= -Z_{\rm ref}^2 \frac{E_1^{(0)}}{n^2}+(Z_{\rm ref} - Z_0)\frac{e^2}{4\pi \epsilon_0}\frac{Z_{\rm ref}}{n^2 a_0}.
\end{aligned}
\label{a_diag}
\end{equation}
Here we have used the results for the energy levels of a hydrogenic state with nuclear charge $Z_{\rm ref}$: $E_n = -Z_{\rm ref}^2 \frac{E_1^{(0)}}{n^2} $, and the 
generalization of the well known 
result\cite{griffiths05} $\langle \phi_n |\frac{1}{r} | \phi_n \rangle = Z_{\rm ref}/(n^2 a_0)$.
}
\begin{figure}[h!]
\begin{center}
\includegraphics[height=0.45\columnwidth]{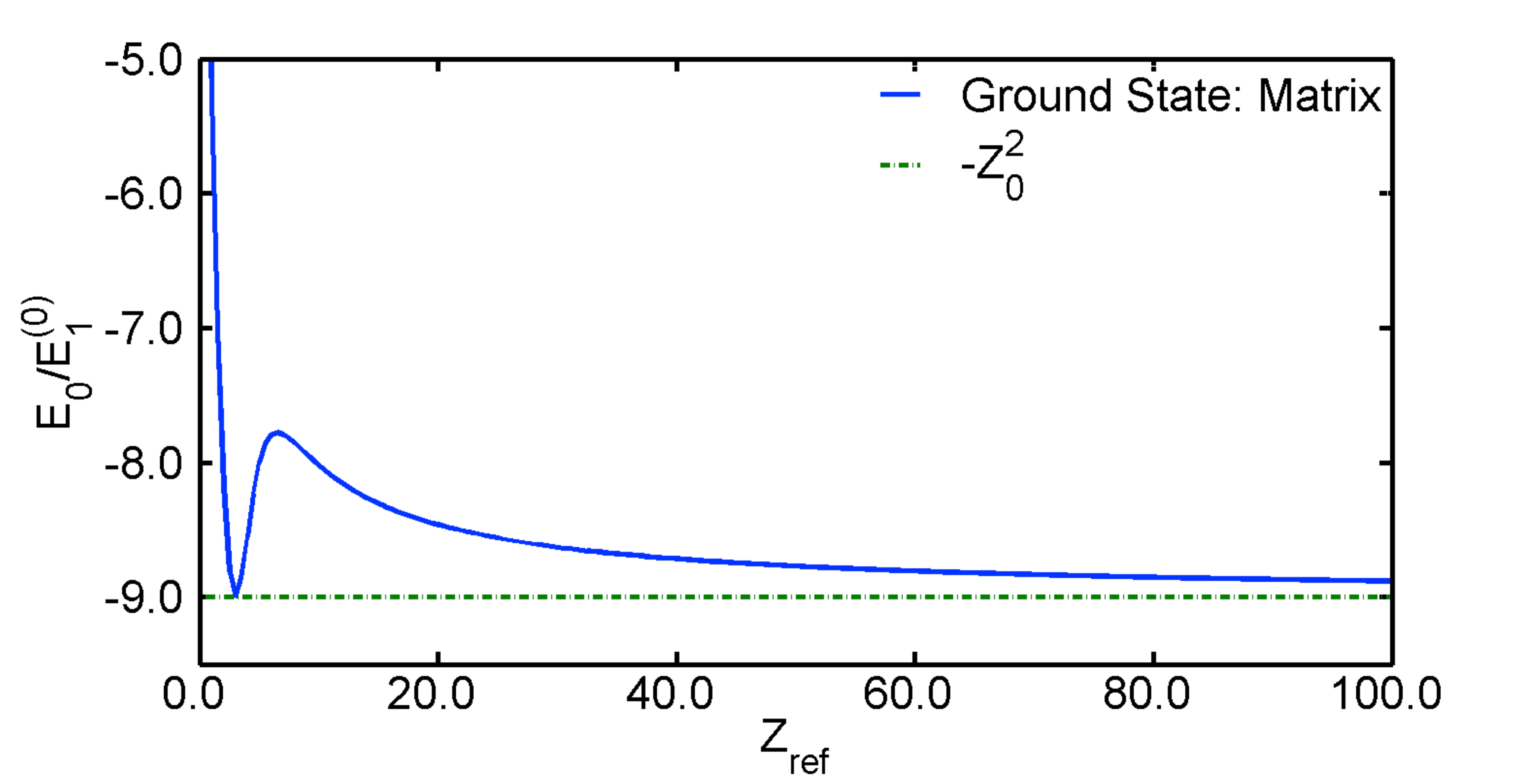} 
\caption{(color online)
\red{The ground state energy for a $Z_0{\rm Hydrogen}$ system with $Z_0 = 3$, vs. $Z_{\rm ref}$. This calculation was performed by
diagonalizing a $18 \times 18$ matrix, and shows that using larger values of $Z_{\rm ref}$ is effective to a degree. Results are
definitely more accurate than using $Z_{\rm ref} = 1$ for example, but the convergence to very accurate results is very slow.}}
\label{fig5}
\end{center}
\end{figure}

\red{Evaluation of the off-diagonal matrix elements proceeds as before; the final result is
\begin{equation}
\langle \phi_m | H | \phi_n \rangle = 2E_1^{(0)} Z_{\rm ref} (Z_{\rm ref} - Z_0)J_{nm},
\label{a_final}
\end{equation}
where
\begin{equation}
\begin{aligned}
J_{nm}  \equiv 
4\frac{(mn)^\frac{1}{2}}{(m+n)^2} ~~&\sum_{i=0}^{m-1}
\frac{(-1)^i}{i!} \frac{(m-1)!}{(m-1-i)!(i+1)!}\biggl(\frac{2n}{m+n}\biggr)^i
\\
& \sum_{j=0}^{n-1}
\frac{(-1)^j}{j!} \frac{(n-1)!}{(n-1-j)!(j+1)!}\biggl(\frac{2m}{m+n}\biggr)^j
(1+i+j)!
\end{aligned}
\label{a_number}
\end{equation}
\noindent is the same number given by Eq.~(\ref{number}) in Section II.}

\red{Now we can simply diagonalize an $N \times N$ matrix for some $Z_{\rm ref}$. In Fig.~(\ref{fig5}) we show the ground state
energy obtained in this way for a targeted system with $Z_0 = 3$. Use of the usual Hydrogen states with $Z_{\rm ref} = 1$
produces a very poor result, $E_0 \approx -5E_1^{(0)}$, compared to the exact result, $E^{\rm ex}_0 = -9E_1^{(0)}$. As we
increase $Z_{\rm ref}$ the result becomes more accurate, and of course exact as $Z_{\rm ref} \rightarrow Z_0$. For larger
values of $Z_{\rm ref}$ the result first deteriorates before steadily improving as $Z_{\rm ref}$ continues to increase towards $100$,
as shown. While the convergence is quite slow, nonetheless this exercise demonstrates that using large values of $Z_{\rm ref}$
allows one to get reasonably accurate results for the reason that the bound state basis set now contains quite a number of
wave functions that describe a particle confined to the origin. Nonetheless, the lack of better accuracy is an indication that a
continuous set of continuum states in the end offers more flexibility than a discrete set of bound states. Note that we did not require
values of $N$ in excess of about $20$ to achieve converged results for all values of $Z_{\rm ref}$ shown.}


\begin{thebibliography}{1}
\bibitem{remark3} For definiteness we write out the definition of the associated Laguerre polynomials that we use in
this paper, following Ref. \citen{griffiths05} as a closed form sum:
\begin{equation}
L_n^{k}(x) = (n+k)!\sum_{j=0}^n {(n+k)! \over (n-j)! (k+j)!} \frac{(-x)^j}{j!}.
\label{laguerre_finite_sum}
\end{equation} 
\noindent Note the additional factor of $(n+k)!$ --- see Ref. \citen{remark2}.
\bibitem{anal} In fact 
\begin{equation}
L_n^k(x) = {(k+1)_n \over n!} M(-n,k+1,x) = {(n+k)! \over k! n!} M(-n,k+1,x),
\label{anal_cont}
\end{equation}
\noindent and it should be clear that this equation is consistent with Eq. (\ref{laguerre_finite_sum}) in Ref. \citen{remark3} and Eq. (\ref{kummer}) in the text.
\end{thebibliography}

\end{document}